\documentclass[aps,prl,groupedaddress,showpacs]{revtex4}
\usepackage{amsmath}
\usepackage{graphicx}

\begin{document}

\title{Absorption and emission of polariton modes in a ZnSe-ZnSSe 
heterostructure}

\author{M. Seemann}
\author{F. Kieseling}
\author{H. Stolz}
\author{M. Florian}
\author{G. Manzke}
\email{guenter.manzke@uni-rostock.de}
\author{K. Henneberger}
\affiliation{University of Rostock, Institute of Physics, 18051 Rostock, Germany}
\author{D. Hommel}
\affiliation{University of Bremen, Institute of Solid State Physics, 28334 Bremen, Germany}

\pacs{71.35.-y, 71.45.Gm, 78.20.-e, 78.67.De}

\begin{abstract}
We investigate the absorption and emission of a $25~nm$ ZnSe layer, 
which was grown on a GaAs buffer and cladded by ZnSSe layers. 
Due to the coupling of light with the exciton resonances polariton
modes propagate through the ZnSe layer. Their interferences appear
as additional peaks in the reflection spectra
and can be explained by the effect of spatial dispersion.
We present additional experimental results for the emission of the sample
after excitation by a pump pulse, showing corresponding 
interference peaks of the polariton modes, whose maxima
strongly decrease to higher energies.
Our exact theoretical analysis shows that the ratio of emission and
absorption is given by the population of the globally defined states of the 
electromagnetic field, i.e. the polariton distribution which is 
generated by the pump pulse. This distribution, being far from thermal 
quasi-equilibrium, shows pronounced peaks at the polariton modes 
depending on the energy of the pump pulse.
\end{abstract}

\maketitle

\section{Introduction} 
In the vicinity of the band edge strong coupling of light with 
exciton resonances leads to a splitting of the light dispersion, known
as polariton effect. Besides the well-known Fabry-Perot modes the 
spatial dispersion leads to an additional series of 
interferences of polariton waves above the exciton resonances 
\cite{NeWu97}.
The behavior of amplitude and phase of the reflected light around these 
resonances was investigated in \cite{SeKi06} for a $25 nm$ ZnSe layer, 
which was grown on a GaAs buffer and cladded by ZnSSe layers.  
Characteristic changes of the phase were found around the polariton 
interferences, which in particular are sensitive to the damping of 
the excitonic resonances. The theoretical description
was based on a broadened 
Elliot formula \cite{Ta95} for the susceptibility of the heavy-hole (hh)
and light-hole (lh) exciton resonances in the active ZnSe layer including
the spatial dispersion. 
Pekar's additional boundary conditions (ABC's) \cite{Pe62} were applied and
according to the microscopic approach in \cite{ScCz04} a weak
penetration of the polarization into the cladding layers was taken into 
account. 
The good agreement of the experimental results for the reflection 
with our theoretical model enables to reconstruct the absorbed intensity 
of a probe pulse in the sample, while the transmitted beam is completely 
absorbed in the GaAs buffer.\\
In this paper we additionally investigate the light emission of the sample at 
low excitation. The experimental results are presented in the
following section. A quantum mechanical and many-body description of
emission was given in \cite{HeKo95,HeKo96} based on Keldysh's technique for
photon Green's functions (GF's). However, in this approach spatial 
dispersion was neglected. A theoretical description of the emission 
including spatial dispersion is a non-trivial problem. 
In \cite{MaMi73,BiMa76} the problem of energy
conservation was discussed in the context of the dielectric approximation 
(DA). 
Recently, starting from energy conservation, one of us (K.H.) \cite{He08} 
presented an exact proof that the ratio between incoherent emission $e$ and 
coherent absorption $a$ is given by the polariton distribution $b$ which for 
quasi-equili\-brium tends towards a Bose distribution . In 
section~\ref{secthe} we give a short outline of this approach and demonstrate 
that the theoretical concept agrees with the experimental findings.

\begin{figure*}[htb]
\includegraphics*[width=\textwidth,height=6cm]{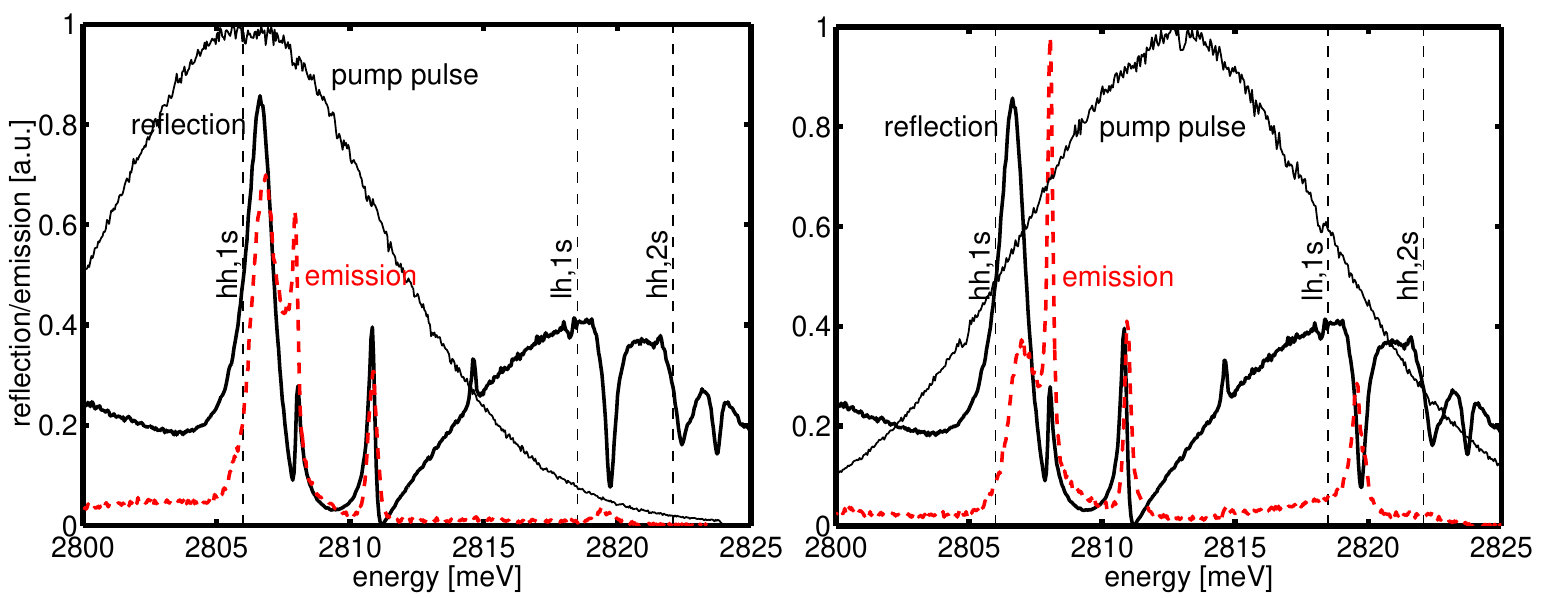}
\caption{Reflection (solid black line) and emission (dashed red) 
for two different pump pulses (thin black).
}\label{fig1}
\end{figure*}

\section{Experimental results}\label{secexp}
Our experiments were performed at a sample grown by molecular beam 
epitaxy and composed of a $25~nm$ ZnSe
layer, cladded by two $1~\mu m$ ${\rm {ZnS}_{x}Se_{1-x}}$ layers with a
sulfur content of $x=7\%$. The sample was contained in a He-flow
cryostat and cooled to a temperature of $T=10~\rm{K}$. 
The experiments were carried out
using $120~\rm{fs}$ (FWHM) pulses with a 
spectral linewidth of $15~\rm{meV}$ (FWHM) centered at different 
photon energies in order to cover the whole spectral domain between 
the heavy- and light-hole exciton resonances of the ZnSe layer. For
details of the measurement of the reflection we refer to \cite{SeKi06}. 
The emission was detected after excitation with pump pulses, which were 
focused on the surface of the sample with a spot diameter of 
$50~\mu m$ under perpendicular incidence.
The emitted light was measured under an angle of $4^o$ by a triple 
additive grating spectrometer with a spectral resolution of 0.1 meV. 
The emission was recorded 
by integrating the spectrometer's CCD over a train of many pump pulses.
The pulse repetition rate is $80 \rm{MHz}$, which corresponds to a 
repetition period of $13~ns$.  \\
The results of our measurements are presented in Fig.~\ref{fig1}.
The spectrum of reflectivity is given by solid black lines in 
both parts of the figure . 
The emitted intensity after pumping with pulses (thin black lines) 
centered at two different energies (left: at the hh-exciton, right: 
between hh- and lh-exciton) is presented with dashed red lines. 
The emission is scaled to be comparable with the reflectivity. 
Additionally, the positions of the exciton resonances are marked by
vertical dashed lines.
Generally, the emission is concentrated on the polariton interference
modes. However, the amplitude of the emission peaks strongly
decreases with increasing energy. 
In \cite{NeWu97} it was found that in dependence of the geometry of the
sample even or odd polariton modes appear more or less pronounced.
As one can see from the figure, this may be different concerning 
reflectivity and emission. While in the reflection the first peak 
directly above the 1s-hh exciton is much higher than the second one, 
in the emission the weight is shifted more to the second peak.  
We demonstrated in our theoretical 
analysis in section~\ref{secthe} that the shape of the emission 
is related to the distribution function of excited polariton states, which 
are generated by the pump pulse. 
\section{Theoretical description}\label{secthe}
\subsection{General balance between absorption and 
emission}\label{subsecgen}
In this section we start with a general description of wave 
propagation through a slab of finite thickness $L$ with the aim
to demonstrate that there is a general connection between emission
and absorption,  summarizing those points of the 
approach in \cite{RiFl08}
which are important for the understanding of the present paper. 
The slab is considered to be infinitely extended in the transverse 
$y$-$z$-direction, and homogeneously excited. For notational 
simplicity, TE-polarized light propagating freely in the transverse 
direction is considered. Due to cylindrical symmetry around the $x$-axis, 
the transverse vector potential 
${\bf A}({\bf r},t)$ can be chosen in the $z$-direction. 
Assuming steady state conditions, the propagation equation 
for the averaged field, after Fourier transforming 
with respect to $\,(y,z) \rightarrow {\bf q}_\perp\,$ and 
$\,t-t' \rightarrow \omega\,$, in this geometry has the structure 
\begin{eqnarray}
\label{1}
\int dx' \, D^{{\rm ret},-1} \, (x,x') \, A(x') \,
= 0&&,
\end{eqnarray}
where $D^{{\rm ret},-1}$ is the inverse of the retarded photon GF
\begin{eqnarray}
\label{2}
D^{{\rm ret},\!-1}(x,\!x'\!)\! 
=\!\left(\frac{\partial^2}{\partial x^2} \!+\! q^{2}_0 \right) \,
\delta (x\!-\!x')\! - \!P^{\rm ret} (x,\!x'\!).
\end{eqnarray}
Here and in what follows, the variables ${\bf q}_\perp$ and $\omega$, which 
enter all equations parametrically only, are omitted where possible. 
In vacuum the propagation in $x$-direction is governed by 
$\, q_0^2 = [(\omega + i\delta)/c)]^{2} - q_{\perp}^{2} \,\,$.
The retarded polarization function of the medium 
$P^{\rm ret}(x,x',q_\perp,\omega)$, acting in Eq.~\ref{2} as 
selfenergy of photons, is related to the susceptibility 
$ \chi $ according to 
\begin{eqnarray}
\label{3}
P^{\rm ret} (x,x') = - \frac{\omega^2}{c^2} \, \chi (x,x') \, .
\end{eqnarray}
The forward propagating 
solution of the homogeneous equation (\ref{1}) has the structure
\begin{eqnarray}
\label{4}
A(x) \, = \, \left\{
\begin{array}{lcl}
e^{i q_0 x} +  r \, e^{-i q_0 x} & \quad \mbox{for} \quad & x < - \frac
{L}{2} \\[0.2cm]
t e^{i q_0 x}  & \mbox{for} & x > \frac{L}{2} \, ,
\end{array}
\right.
\end{eqnarray}
where $\, r,t \,$ are reflection and transmission coefficients for the field 
amplitudes. Due to the symmetry  $x \leftrightarrow -x$, $A(-x)$ is a 
solution, too (backward propagating, i.e., incidence from right). It
is the advantage of this approach that it works in general, without
specifying the solutions inside the slab. Of course, these solutions
have to be determined to calculate the coefficients $t$, and $r$. This
will be done in section~\ref{subsecspe}.  
Following Poynting's theorem, the $x$-component of the Poynting vector 
in the considered geometry is related to the absorbed field energy 
$W=j\cdot E$ by 
\begin{eqnarray}
\label{5}
S\left(\frac{L}{2},\omega \right)-S\left(-\frac{L}{2},\omega 
\right)=-\int\limits_{-L/2}^{L/2} dx \, 
W(x,\omega).
\end{eqnarray}
The current density is related via $j=\partial P/\partial t$ to the
polarization $P=\chi E$, and the electric field via 
$E=-\partial A/\partial t$ to the vector potential $A$. 
Assuming a monochromatic wave of frequency $\omega_0$ incident 
in the $(q_0 , {\bf q}_{\perp,0})$-direction 
\begin{eqnarray}
\label{6}
 A (x, {\bf q}_\perp , \omega) = \frac{1}{2} [A_0 (x, {\bf q}_{\perp , 0} , 
\omega_0) \delta (\omega - \omega_0) 
\delta_{  {\bf q}_\perp ,{\bf q}_{\perp , 0} } \\ \nonumber 
+ A_0^* (x, {\bf q}_{\perp 
, 0} , \omega_0) \delta (\omega + \omega_0) 
\delta_{  {\bf q}_\perp ,-{\bf q}_{\perp , 0} }] \, , 
\end{eqnarray}
in Eq.~(\ref{5}) after straightforward calculation for each 
$\, \omega_0 \rightarrow \omega \,$ and ${\bf q}_{\perp,0}  \rightarrow 
{\bf q}_{\perp} $ yields
\begin{eqnarray}
\label{7}
1-|r|^2 - |t|^2 = a= \frac{i}{2q_0}\hat{\cal{P}} \, ,
\end{eqnarray}

\vspace*{-3mm}
\begin{eqnarray} 
\label{8}
\hat{\cal{P}} = \int dx dx' A^{*}(x) \, \hat{P}(x,x')\,A(x').
\end{eqnarray}
Eq. (\ref{7}) simply balances the 
incoming intensity $(\sim 1)$ into absorption $a$ and the sum of the reflected 
and transmitted intensity. On the other hand, due to the well-known GF 
identities 
\protect\cite{Ke65} 
\begin{eqnarray} 
\label{9}
\hat{P} =  P^{\rm ret} -  P^{\rm adv} = P^> - P^< \,\, ,
\end{eqnarray}
the absorption balances generation $ i P^>(x,x')$ and recombination 
$  i P^<(x,x')$ of excitons in the medium. \\
The incoherent or correlated emission is 
defined as the one without external sources,  i.e. for vanishing averaged 
fields. In this case, due to the non-commuting field operators, 
the symmetrized Poynting vector 
$\,\,{\bf S} = \frac{1}{2 \mu_0} \, ({\bf E} \times {\bf B} - {\bf B} \times 
{\bf E})\,\,$ and absorption/emission  
$\,\,W = \frac{1}{2} \, ({\bf j}{\bf E} + {\bf E} {\bf j})$, 
respectively, have to be used. 
Following the approach in Ref.~\protect\cite {HeKo95}
the intensity of the emitted light is given by 
\begin{eqnarray}
\label{10}
S(x) = \frac{1}{2} \left\{ \frac{\partial}{\partial x'} \, 
\left[ D^{>} (x,x') + D^{<} (x,x') \right] \right\}_{x'=x} 
\, ,
\end{eqnarray}
and the emitted field energy by
\begin{eqnarray} 
\label{10a}
W(x)\!=\!\!\int\!\!dx'\!\!\left[P^{>}\!(x\!,\!x'\!)D^{<}\!(x'\!,\!x)\!-\!
P^{<}\!(x\!,\!x'\!) D^{>}\!(x'\!,\!x) \right]\!.
\end{eqnarray} 
$ D^{\stackrel{>}{<}}(x,x', {\bf q}_\perp , \omega) \,\,$
and $ P^{\stackrel{>}{<}}(x,x', {\bf q}_\perp , \omega) \,\,$ are the Keldysh 
components of the photon GF and of the polarization function 
(see Eq.~(\ref{9})), respectively. 
Here as in Eqn's~(\ref{1}-\ref{9}) the variables ${\bf q}_\perp\,$ and 
$\omega$ are introduced by Fourier transform, but are dropped, since 
they only appear as parameters. \\
The polarization function 
\begin{eqnarray}
\label{11}
P^{\stackrel{>}{<}}(x,x') = P^{\stackrel{>}{<}}_m (x,x')
\mp  i \delta \, \frac{2 \omega}{c^{2}} \Theta (\pm \omega ) \, \delta (x-x')
\end{eqnarray}
comprises a medium part $P_m$ and a part which describes the vacuum 
as an infinitely weak ($\delta\rightarrow0$) absorber \protect\cite{HeKo96}.  
As shown in \protect\cite{RiFl08}, the formal solution (optical theorem) of 
the photon Dyson equation can be used in (\ref{10a}) yielding after 
straightforward calculation the incoherent absorption as
\begin{eqnarray}
\label{12}  
{\cal W} = \int dx\,W(x) = - \frac{i}{q_0} {\cal P}^< , 
\end{eqnarray}
As usual,
a distribution function $ b $ will be attributed by definition
to the global recombination ${\cal P}^<(\omega, {\bf q}_\perp)$ 
defined in (\ref{12}) 

\vspace*{-5mm}
\begin{eqnarray}
\label{13}
{\cal P}^< = \int dx \, dx' A^{*}(x)  \, P^<_m (x,x') \, A (x')
=\,b\,\hat{\cal{P}} . 
\end{eqnarray} 

\vspace*{-1mm}
\noindent
Now Poynting's theorem (\ref{5}) provides the incoherent 
energy flow $ e=S(L/2) = -S(-L/2) $ from the incoherent absorption 
(\ref{12}): $ 2e = - {\cal W} $. Then from (\ref{8}), (\ref{12}), and 
(\ref{13}), for any 
frequency $\, \omega > 0 \,$ and propagation direction 
$\,\, {\bf q} = (q_0 , {\bf q}_\perp) \,$, we obtain for the spectrally and 
directionally resolved energy flow (emission)
\begin{eqnarray}
\label{21}  
e = \frac{i}{2q_0} {\cal P}^< = b \,a \, .
\end{eqnarray} 
The distribution $b$ is 
accessible to direct observation in experiments measuring the 
(incoherent) emission and the classical absorption, i.e., transmission and 
reflection. It generalizes Planck's formula for 
the black body radiation to the non-equilibrium radiation of an excited 
medium in the steady state. Relation (\ref{21}) is generally valid. 
It was derived without specifying the fields inside the slab. In this 
respect it works as a criterion which has to be fulfilled for each model 
applied to the polariton pulse propagation, in particular for all applied 
ABC's in the calculations. In contrast to the authors in \cite{MaMi73,BiMa76},
we find energy conservation valid in the DA, too \cite{RiFl08}.

For quasi-equilibrium, due to the Kubo-Martin-Schwin\-ger 
condition \protect\cite {KMS57}, 
the distribution $b(\omega, {\bf q}_\perp)$ 
develops into a Bose distribution $\,b (\omega) = \left( {\rm exp} 
\left[\beta(\hbar \omega - \mu)\right] - 1 \right)^{-1}\,$ with the
temperature $T$ ($\beta=1/kT$) and the chemical potential $\mu$, 
being independent of $ {\bf q}_\perp $.  
Measuring emission and absorption enables to check whether 
quasi-equilibrium is realized. 
\subsection{Absorption and Emission of the investigated 
sample }\label{subsecspe}
\begin{figure*}[htb]
\includegraphics*[width=\textwidth,height=6cm]{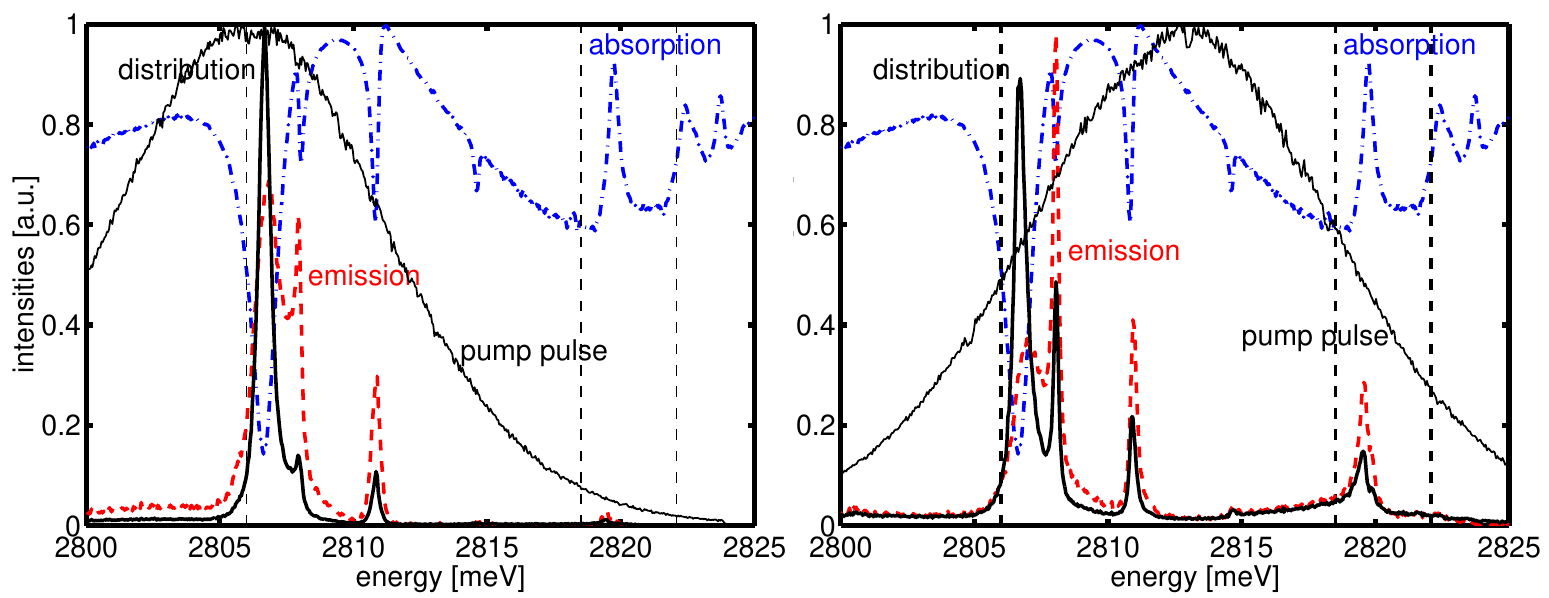}
\caption{Absorption (dash-dotted blue line), emission (dashed red) 
and distribution function (solid black) for two different 
pump pulses (thin black).
}\label{fig2}
\end{figure*}
In our sample absorption and emission is investigated in the vicinity
of the heavy-hole (hh) and light-hole (lh) exciton resonances of the 
$25~nm$ ZnSe layer which is cladded by two $1~\mu m$ ZnSSe layers. As
we have demonstrated in \cite{SeKi06}, the reflection is dominated by
polariton interferences (see Fig.~\ref{fig1},~\ref{fig2}) and can be described 
considering a dielectric function for the active ZnSe-layer, which
contains the 1s-hh, 2s-hh, and
1s-lh exciton resonances including their spatial dispersion. The
polariton dispersion splits into four branches and additional boundary
conditions (ABC's) for the single polariton waves at the surfaces have to be
considered. The experiments fit well applying Pekar's ABC's and
assuming a small increased effective ZnSe layer ($25.5~nm$) which takes
into account the penetration of the excitonic wave function into the
cladding layers. 
This treatment is a simple way to circumvent the 
costly numerical microscopic calculation of the spatially resolved 
polarization presented in \cite{ScCz04} and the approach in \cite{MuZi02}. 
The experimental results could not be verified applying Ting's ABC's 
\cite{TiFr75}. However, the dielectric approximation \cite{MaMi73,BiMa76},
being a combination of differently weighted ABC's of Ting and Pekar, 
gives good agreement, since the weight is strongly focused to Pekar's 
ABC's. \\
The good agreement of our
theoretical calculations with the experimental results in Fig.~\ref{fig1}
(see also \cite{SeKi06}) encourages us to calculate the absorbed intensity
of light according to Eq. (\ref{7}). In contrast to the single slab considered
in the preceeding section \ref{subsecgen} no light passes through 
the back side of the sample, since it is completely absorbed in the GaAs 
buffer layer. That is to say, there is no total transmission, and 
the absorption in Eq.~(\ref{7}) is simply $a=1-R$, where $R$ is the 
intensity of the reflected light normalized to the intensity of the 
incoming light.
The total absorption of the 
sample is presented in Fig.~\ref{fig2} by the dash-dotted blue lines.
The pump pulses and the emission correspond to those in Fig.~\ref{fig1}.
Additionally, the quotient of emission and absorption is given 
(solid black line), which represents the distribution function $b$
of polariton states (\ref{21}) in the sample.
It shows clearly that the polariton modes are predominantly occupied.
Depending of the position of the pump pulse, the occupation of the 
single polariton modes varies.
This behavior indicates that the detected emission comes rather from the 
excited polariton states, generated by the pump pulse, than from a 
quasi-equilibrium distribution.

\begin{acknowledgments}
We would like to thank the Deutsche Forschungsgemeinschaft for support through 
the Sonder\-for\-schungs\-bereich 652.  
\end{acknowledgments}



\begin{thebibliography}{[1]}

\bibitem{NeWu97} U. Neukirch, K. Wundke, Phys. Rev. {\textbf B 55}\,,\,
  15408 (1997).
  
\bibitem{SeKi06} M. Seemann, F. Kieseling, H. Stolz, 
  G. Manzke, and K. Henneberger,
  Solid State Comm. {\textbf 138}\,,\, 457\, (2006).

\bibitem{Ta95} C. Tanguy, Phys. Rev. Lett., {\textbf 75}, 4090 (1995).
  
\bibitem{Pe62} S. J. Pekar, Sov. Phys. Sol. State {\textbf 4}, 953 (1962).
  
\bibitem{ScCz04} S. Schumacher, G. Czycholl, F. Jahnke, I. Kudyk, H.
  I. R\"uckmann, J. Gutowski, A. Gust, G. Alexe, and D. Hommel, Phys.
  Rev. {\textbf B 70}, 235340 (2004).
  
\bibitem{MuZi02}
 E.A. Muljarov and R. Zimmermann, Phys. Rev. B {\bf 66}, 235319 (2002).

\bibitem{HeKo95}
 K. Henneberger and S. W. Koch,
 in: Microscopic Theory of Semiconductors: Quantum Kinetics,
 Confinement and Lasers, edited by  S. W. Koch 
 (World Scientific Publ. Co. Pte. Ltd. 1995).

\bibitem{HeKo96}
 K. Henneberger and S. W. Koch,
 Phys. Rev. Lett. {\bf 76}, 1820 (1996).

\bibitem{MaMi73}
A. A. Maradudin and D.L. Mills, Phys. Rev. B {\bf 7}, 2787 (1973).

\bibitem{BiMa76} M. F. Bishop, and A. A. Maradudin,
  Phys. Rev. {\textbf 14}, 3384 (1976).

\bibitem{He08}
K. Henneberger, arXiv:0710.5686 [condmat.str-l]

\bibitem{Pe57}
S.I. Pekar, Zh. Eksp. Teor. Fiz. {\bf 33}, 1022 (1957) [Soviet Phys. JETP {\bf 6}, 785 (1958)].

\bibitem{Ke65}
L. V. Keldysh, Zh. Eksp. Teor. Fiz. {\bf 47}, 1515 (1964); [Sov. Phys. JETP {\bf 20, 1018 (1965)}].

\bibitem{RiFl08}
F. Richter, M. Florian, and K. Henneberger, 
Europhys. Lett. {\bf 81}\,,\, 67005\,(2008). 

\bibitem{KMS57}
R. Kubo, J. Phys. Soc. Japan {\bf 12}, 570 (1957)

\bibitem{TiFr75}
C. S. Ting, M. J. Frankel, and J. L. Birman, Solid State Comm. 
{\textbf 17}\,,\, 1285\, (1975).

\end{thebibliography}
\end{document}